\documentclass[%
 reprint,
 amsmath,amssymb,
 aps,
]{revtex4-1}

\usepackage{graphicx}
\usepackage{dcolumn}
\usepackage{bm}
\usepackage[colorlinks=true,citecolor=blue,linkcolor=blue,urlcolor=blue]{hyperref}

\def\be{\begin{equation}}
\def\ee{\end{equation}}

\begin{document}

\preprint{APS/123-QED}

\title{Dark Energy from Gauss-Bonnet and non-minimal couplings}

\author{L. N. Granda}
 \email{luis.granda@correounivalle.edu.co} 
\author{D. F. Jimenez}%
 \email{jimenez.diego@correounivalle.edu.co}
\affiliation{%
Departamento de Fisica, Universidad del Valle
\\
A.A. 25360, Cali, Colombia 
}%


%
%

\begin{abstract}
We consider a scalar-tensor model of dark energy with Gauss-Bonnet and non-minimal couplings. Exact cosmological solutions were found in absence of potential, that give  equations of state of dark energy consistent with current observational constraints, but with different asymptotic behaviors depending on the couplings of the model. 
A detailed reconstruction procedure is given for the scalar potential and the Gauss-Bonnet coupling
for any given cosmological scenario. Particularly, we consider conditions for the existence  of a variety of cosmological solutions with accelerated expansion, including quintessence, phantom, de Sitter, Little Rip. For the case of quintessence and phantom we have found a scalar potential of the Albrecht-Skordis type, where the 
potential is an exponential with a polynomial factor.
\begin{description}
\item[PACS numbers]
98.80.-k, 95.36.+x, 04.50.Kd
\end{description}
\end{abstract}

\maketitle


\section{\label{intro}introduction}
A significant number of extensions of general relativity modifying the behavior of Einstein gravity at cosmic scales,  have been introduced lately in order to explain the observed accelerated expansion of the universe \cite{perlmutter}-\cite{planck2}. These models range from modifications of the source terms that involve scalar fields of different nature like quintessence, phantom, tachyion, dilaton (see \cite{copeland}-\cite{sergeiod} for review), including different couplings of scalar field to curvature (scalar-tensor theories) \cite{chiba}-\cite{granda3}, to modifications of the geometrical terms involving curvature, known as modified gravity theories \cite{capozziello1}-\cite{ tsujikawa}.\\
Among all possible explanations for the observed cosmic acceleration  the scalar-tensor theories are widely considered since the couplings of scalar fields to curvature naturally appear in the process of quantization on curved space time \cite{ford, birrel}, from compactifications of higher dimensional gravity theories \cite{ lamendola1},  in the next to leading order corrections in the $\alpha'$ expansion of the string theory \cite{metsaev, cartier, green}.
Perhaps the most studied and the  simplest one of these couplings is the non-minimal coupling of the type $\xi R\phi^2$. The role of this coupling in the dark energy (DE) problem has been studied in different works, including the constraint on $\xi$ by solar system experiments \cite{chiba}, the existence and stability of cosmological scaling solutions \cite{uzan,lamendola}, perturbative aspects and incidence on CMB \cite{perrotta,riazuelo}, tracker solutions \cite{perrotta1}, observational constraints and reconstruction \cite{boisseau, polarski, capozziello3}, the coincidence problem \cite{tchiba}, super acceleration and phantom behavior \cite{vfaraoni}-\cite{polarski1}, and asymptotic de Sitter attractors \cite{vfaraoni1}. 
Another interesting interaction that gives good results in the study of the dark energy is the coupling between the scalar field and the Gauss-Bonnet (GB) invariant . Although the GB term is topologically invariant in four dimensions, nevertheless it affects the cosmological dynamics when it is coupled to a scalar field, contributing second order differential terms to the equations of motion. This coupling has been proposed to address the dark energy problem in \cite{sergei4}, showing attractive features like preventing the Big Rip singularity in phantom cosmology. Different  aspects of accelerating cosmologies with GB correction have been discussed in \cite{sergei5}-\cite{capozziello5}. All these studies demonstrate that it is quite plausible that the scalar-tensor couplings predicted by fundamental theories may become important at current, low-curvature universe.\\
In this paper we consider a scalar-tensor model that contains the above mentioned two couplings:  the non-minimal coupling to the curvature and to the Gauss-Bonnet invariant, and analyze  the role of these couplings in the late time universe.  The understanding of the effect of these non-minimal couplings could provide clues about how the fundamental theories at high energies manifest at cosmological scales. We consider two different approaches: first we find exact cosmological solutions in absence of potential, that give consistent description of different DE scenarios. It is shown that depending on values of the couplings, the universe can be currently undergoing a transition from quintessence to phantom phase or from phantom to quintessence phase. 
In the second approach we reconstruct the potential and the GB coupling for various interesting cosmological scenarios beginning with power-law expansion and considering also Big Rip and Little Rip solutions. For power-law solutions we found a potential that was considered before in connection with low energy limit of M-theory \cite{andreas, skordis}.
The paper is organized as follows. In Sec. II we introduce the model of the scalar field with  non-minimal and GB couplings and give a detailed reconstruction scheme for a given cosmological scenario. In section III we present exact solutions of the model with zero potential and study the behavior of the corresponding equations of state. In sec. IV we present some samples of late time cosmologies and make reconstruction of the scalar potential and GB coupling. Sec. V is dedicated to some summary and discussion. 
\section{The scalar-tensor model and field equations}
Let's start with the action for the scalar field where in addition to the GB coupling we consider the non-minimal coupling of the scalar field to curvature. As will be seen below, the non-linear character of the cosmological equations  makes the integration of the same ones very difficult for a given set of initial conditions. Nevertheless given that we know in advance some cosmological scenarios that represent probable states of expansion of the universe, we  can deal with the inverse problem: for a given cosmological solution try the reconstruction of the model. Below we give a  general approach to reconstruct the scalar-tensor model with non-minimal and GB couplings for any given cosmological scenario \cite{sergei5}. The action for the model with scalar field and matter is given by

\begin{eqnarray}\label{eq1}
S_{\phi}=&&\int d^{4}x\sqrt{-g}\Bigg[\frac{1}{2 {\kappa}^2 }R-\frac{1}{2}\partial_{\mu}\phi\partial^{\mu}\phi-\frac{1}{2}\xi R{\phi}^{2}\nonumber\\
&&-V(\phi)-\eta(\phi){\cal{G}}+{\cal{L}}_m\Bigg],
\end{eqnarray}
where ${\kappa}^2 =8\pi G$,  ${\cal{G}}=R^2-4R_{\mu\nu}R^{\mu\nu}+R_{\mu\nu\lambda\rho}R^{\mu\nu\lambda\rho}$ is the GB invariant, and ${\cal{L}}_m$ is the Lagrangian of perfect fluid with energy density ${\rho _m}$ and pressure $ {P_m}$. We will consider the spatially-flat Friedmann-Robertson-Walker (FRW) metric. 

\begin{equation}\label{eq2}
ds^2=-dt^2+a(t)^2\sum_{i=1}^{3}(dx_i)^2
\end{equation}
The Friedman equations with Hubble parameter $H=\dot{a}/a$ are
\begin{eqnarray}\label{eq3}
  H^2 = &&\frac{\kappa^2}{3} \Bigg(\frac{1}{2} \dot{\phi}^2+V(\phi) + 6 \xi H \dot{\phi} \phi +
  3\xi H^2 \phi^2\nonumber\\
&&+24H^3\frac{d\eta}{d\phi}\dot{\phi}+{\rho_m }\Bigg)
\end{eqnarray}

\begin{eqnarray}\label{eq4}
- 2\dot H - 3{H^2} = &&{\kappa ^2}\Bigg[\frac{1}{2}{{\dot \phi }^2} - V(\phi ) - 2\xi \left( {{{\dot \phi }^2} + \phi \ddot \phi } \right)\nonumber\\
&& - 4\xi H\phi \dot \phi - \xi \left( {2\dot H + 3{H^2}} \right){\phi ^2} - 8{H^2}\frac{{d\eta }}{{d\phi }}\ddot \phi \nonumber\\
&&- 8{H^2}\frac{{{d^2}\eta }}{{d{\phi ^2}}}{{\dot \phi }^2} - 16H\dot H\frac{{d\eta }}{{d\phi }}\dot \phi - 16{H^3}\frac{{d\eta }}{{d\phi }}\dot \phi\nonumber\\
&&+{P_m}\Bigg]
\end{eqnarray}
Deleting the potential from eqs. (\ref{eq3}) and (\ref{eq4}) one finds
\begin{eqnarray}\label{eq5}
{\rho _m}+ {P_m}=&&-2\xi H\phi \dot \phi - 8{H^3}\frac{{d\eta }}{{d\phi }}\dot \phi - {{\dot \phi }^2} + 2\xi {{\dot \phi }^2}\nonumber\\
&& + 8{H^2}\frac{{{d^2}\eta }}{{d{\phi ^2}}}{{\dot \phi }^2} + 2\xi \phi \ddot \phi + 8{H^2}\frac{{d\eta }}{{d\phi }}\ddot \phi \nonumber\\
&& - \frac{2}{{{\kappa ^2}}}\dot H + 2\xi {\phi ^2}\dot H + 16H\dot H\frac{{d\eta }}{{d\phi }}\dot \phi
\end{eqnarray}
which can be written in the form

\begin{eqnarray}\label{eq6}
\frac{1}{2}({\rho _m}+ {P_m})=&&\xi\phi\ddot{\phi}-\xi H\phi\dot{\phi}-(\frac{1}{2}-\xi)\dot{\phi}^2-(\frac{1}{\kappa^2}-\xi\phi^2)\dot{H}\nonumber\\
&&+4\left(H^2\frac{d^2\eta}{dt^2}+2H\dot{H}\frac{d\eta}{dt}-H^3\frac{d\eta}{dt}\right)
\end{eqnarray}
If additionally we have in mind that

\begin{equation}\label{eq7}
a\frac{d}{dt}\left[\frac{H^2}{a}\frac{d\eta}{dt}\right]=H^2\frac{d^2\eta}{dt^2}+2H\dot{H}\frac{d\eta}{dt}-H^3\frac{d\eta}{dt}
\end{equation}
then, the eq. (\ref{eq5}) can be written as

\begin{eqnarray}\label{eq8}
4a\frac{d}{dt}\left[\frac{H^2}{a}\frac{d\eta}{dt}\right]=&&\frac{1}{2}({\rho _m}+ {P_m})+(\frac{1}{2}-\xi)\dot{\phi}^2+(\frac{1}{\kappa^2}-\xi\phi^2)\dot{H}\nonumber\\
&&-\xi\phi\ddot{\phi}+\xi H\phi\dot{\phi}
\end{eqnarray}
Let's define the following function

\begin{eqnarray}\label{eq9}
R(t)=&&\frac{1}{2}({\rho _m}+ {P_m})+(\frac{1}{2}-\xi)\dot{\phi}^2+(\frac{1}{\kappa^2}-\xi\phi^2)\dot{H}-\xi\phi\ddot{\phi}\nonumber\\
&&+\xi H\phi\dot{\phi}
\end{eqnarray}
which allows to rewrite the Eq. (\ref{eq8}) in the compact form:

\begin{equation}\label{eq10}
R(t)=4a\frac{d}{dt}\left[\frac{H^2}{a}\frac{d\eta}{dt}\right]
\end{equation}
This equation can be solved with respect to $\eta$ as

\begin{equation}\label{eq11}
\eta(t)=\frac{1}{4}\int^{t}dt_1\frac{a(t_1)}{H^2(t_1)}\int^{t_1}dt_2\frac{R(t_2)}{a(t_2)}
\end{equation}
From (\ref{eq3}) one can write the potential as

\begin{equation}\label{eq12}
V(t)=S(t)-24H^3\frac{d\eta}{dt}
\end{equation}
where the function $S(t)$ is defined as

\begin{equation}\label{eq13}
S(t)=\frac{3}{\kappa^2}H^2-3\xi H^2\phi^2-6\xi H\phi\dot{\phi}-\frac{1}{2}\dot{\phi}^2-{\rho _m}
\end{equation}
and using the Eq. (\ref{eq11}) one finds

\begin{equation}\label{eq14}
V(t)=S(t)-6a(t)H(t)\int^{t}dt_1\frac{R(t_1)}{a(t_1)}
\end{equation}
The Eqs. (\ref{eq11}) and (\ref{eq14}) allow to reconstruct the model for a given scalar field $\phi(t)$ and Hubble function $H(t)$.
In order to express the potential $V$ and the Gauss-Bonnet coupling $\eta$ as functions of the scalar field, we can assume that 

\begin{eqnarray}\label{eq15}
\begin{aligned}
&t=f(\phi)\\
& a={a_0}e^{g(t)},\,\,\,\,\ H=g'(t)
\end{aligned}
\end{eqnarray}
where $f$ y $g$ are given functions. Replacing (\ref{eq15}) in  (\ref{eq11}) and (\ref{eq14}) we find

\begin{eqnarray}\label{eq16}
\eta(\phi)=&&\frac{1}{4}\int^{\phi}d\phi_1 f'(\phi_1)\frac{e^{g(f(\phi_1))}}{[g'(f(\phi_1))]^2}\nonumber\\
&&\times\int^{\phi_1}d\phi_2 f'(\phi_2)e^{-g(f(\phi_2))}R(f(\phi_2))
\end{eqnarray}
\begin{eqnarray}\label{eq17}
V(\phi)=&&S(f(\phi))-6g'(f(\phi))e^{g(f(\phi))}\int^{\phi}d\phi_1f'(\phi_1)\nonumber\\
&&\times e^{-g(f(\phi_1))}R(f(\phi_1))
\end{eqnarray}
These equations allow to reconstruct the model by given  $f$ y $g$. In order to have explicit solutions for $R(f(\phi ))$  and $S(f(\phi ))$ we need to take into account the conservation equation for the matter component with constant equation of state $w$
\[{{\dot \rho }_m} + 3H(1 + \omega ){\rho _m} = 0\,\,\,\,\,\,\, \Rightarrow \,\,\,\,\,\,{\rho _m} = {\rho _{m0}}{a^{ - 3(1 + \omega )}}\]
From $a =a_0 {e^{g(t)}}$, follows
\[{\rho _m} = {g_0}{e^{ - 3(1 + \omega )g(t)}}\,\,\,\,\,\,\,\,\,{\text{where  }}\,\,\,\,\,\,\,\,\,{g_0} = {\rho _{m0}}a_0^{ - 3(1 + \omega )}\]
This way it is obtained that

\begin{eqnarray}\label{eq18}
R(f(\phi )) =&& \frac{{\xi \phi g'(f(\phi ))}}{{f'(\phi )}} - \frac{\xi }{{f'{{(\phi )}^2}}} + \frac{1}{{2f'{{(\phi )}^2}}}+ \frac{{\xi \phi f''(\phi )}}{{f'{{(\phi )}^3}}}\nonumber\\
&& + \frac{{g''(f(\phi ))}}{{{\kappa ^2}}}+ \frac{1}{2}{g_0}(\omega  + 1){e^{ - 3(\omega  + 1)g(f(\phi ))}} \nonumber\\
&& - \xi {\phi ^2}g''(f(\phi ))
\end{eqnarray}
\begin{eqnarray}\label{eq19}
S(f(\phi )) =&&  - \frac{{6\xi \phi g'(f(\phi ))}}{{f'(\phi )}} - \frac{1}{{2f'{{(\phi )}^2}}} + \frac{{3g'{{(f(\phi ))}^2}}}{{{\kappa ^2}}}\nonumber\\
&& - 3\xi {\phi ^2}g'{(f(\phi ))^2} - {g_0}{e^{ - 3(\omega  + 1)g(f(\phi ))}}
\end{eqnarray}
The Eqs. (\ref{eq18}) and (\ref{eq19}) after being replaced in the Eqs. (\ref{eq16}) and (\ref{eq17}) give the reconstructed model by known functions $f$ and $g$. Therefore any cosmological scenario for the model (\ref{eq1}) encoded in the functions $f(\phi)$ and $g(t)$ with the GB coupling and potential given by (\ref{eq16}) and (\ref{eq17}) can be realized. In section IV we provide some important examples, including cosmologies with phase of super-acceleration. In the next section we consider the model without potential, find exact solutions and analyze possible late time cosmological scenarios.
\section{The dynamics without potential. Exact solutions}
As we will show, the model contains interesting solutions even in absence of the scalar potential. 
To this end, we write down the equation of motion for the scalar field

\be\label{eq10a}
\ddot{\phi}+3H\dot{\phi}+\frac{dV}{d\phi}+6\xi (2H^2+\dot{H})\phi+24H^2\left(H^2+\dot{H}\right)\frac{d\eta}{d\phi}
\ee
Setting $V=0$ and turning to the e-folding variable $x=\ln a$ we can write the Eqs.  (\ref{eq3}) and (\ref{eq10a}) in the form
\begin{eqnarray}\label{eq10b}
H^2=&&\frac{\kappa^2}{3}\Big[\frac{1}{2}H^2\left(\frac{d\phi}{dx}\right)^2+3\xi H^2\left(2\phi\frac{d\phi}{dx}+\phi^2\right)\nonumber\\ &&+24H^4\frac{d\eta}{dx}\Big]
\end{eqnarray}
\begin{eqnarray}\label{eq10c}
\frac{d}{dx}\left[H^2\left(\frac{d\phi}{dx}\right)^2\right] &&+6H^2\left(\frac{d\phi}{dx}\right)^2+6\xi\left(4H^2+\frac{dH^2}{dx}\right)\phi\frac{d\phi}{dx}\nonumber\\ &&+24H^2\left(2H^2+\frac{dH^2}{dx}\right)\frac{d\eta}{dx}=0
\end{eqnarray}

In order to integrate these equations we  consider the following expression for the GB coupling
\be\label{eq10d}
\frac{d\eta}{dx}=\frac{1}{H^2}\left(g_1+g_2\phi^2\right)
\ee
By using this expression for the GB coupling, the Eqs. (\ref{eq10b}) and (\ref{eq10c}) reduce to the following
\be\label{eq10e}
\left(\frac{d\phi}{dx}\right)^2+12\xi \phi \frac{d\phi}{dx}+6\left(\xi+8g_2\right)\phi^2+48g_1-\frac{6}{\kappa^2}=0
\ee
and
\be\label{eq10f}
\begin{aligned}
&\left[\frac{d}{dx}\left(\frac{d\phi}{dx}\right)^2+6\left(\frac{d\phi}{dx}\right)^2+24\xi\phi\frac{d\phi}{dx}+48\left(g_1+g_2\phi^2\right)\right]H^2\\&+\left[\left(\frac{d\phi}{dx}\right)^2+6\xi\phi\frac{d\phi}{dx}+24\left(g_1+g_2\phi^2\right)\right]\frac{dH^2}{dx}=0
\end{aligned}
\ee
by restricting the constant $g_1$ to the value $g_1=1/(8\kappa^2)$, the last term in (\ref{eq10e}) disappears, reducing it to
\be\label{eq10g}
\frac{d\phi}{dx}=\left(-6\xi\pm\sqrt{6\xi(6\xi-1)-48g_2}\right)\phi,
\ee
which gives the solution
\be\label{eq10h}
\phi(x)=\phi_0 e^{-\lambda x},\;\;\;\;\, \lambda=6\xi\pm\sqrt{6\xi(6\xi-1)-48g_2}
\ee
Replacing this solution into the equation of motion (\ref{eq10f}) leads to the following equation for the Hubble parameter
\be\label{eq10i}
\begin{aligned}
&\left(\lambda^2-6\xi\lambda+24g_2+\frac{3}{\kappa^2\phi_0^2}e^{2\lambda x}\right)\frac{dH^2}{dx}+\\& 2\left(-\lambda^3+3\lambda^2-12\xi\lambda+24g_2+\frac{3}{\kappa^2\phi_0^2}e^{2\lambda x}\right)H^2=0
\end{aligned}
\ee
After solving this equation we can find the dark energy equation of state (EOS) as 
\be\label{eq10j}
w_{DE}=-1-\frac{1}{3H^2}\frac{dH^2}{dx}
\ee
In order to integrate the equation (\ref{eq10i}) we can assume some simplifications. First we can note that at far future ($x>>1$) if $\lambda>0$, the exponential terms will dominate over the constants and the solution of the equation reduces to
\be
H^2=H_0^2 e^{-2x}
\ee
where $H_0$ is the integration constant. The EOS for this solution is the constant $w=-1/3$, which corresponds to the divide between the decelerated and accelerated expansion. This means that (provided $\lambda>0$) even if currently the  EOS is close to $w_0=-1$, then at far future the dark energy EOS tends to $w=-1/3$. On the other hand, if $\lambda<0$ then at far future the exponential terms in the Eq. (\ref{eq10i}) become subdominant and the solution to the equation will depend on the constant terms and reduces to
\be
H^2=H_0^2 \exp\left(\frac{2\lambda^3-6\lambda^2+24\xi\lambda-48g_2}{\lambda^2-6\xi\lambda+24g_2}x\right)
\ee
which gives the constant EOS
\be\label{eq10k}
w=-1-\frac{1}{3}\left(\frac{2\lambda^3-6\lambda^2+24\xi\lambda-48g_2}{\lambda^2-6\xi\lambda+24g_2}\right)
\ee
Thus depending on the values of $\xi$ and $g_2$, at far future the EOS tends asymptotically to the cosmological constant if $2\lambda^3-6\lambda^2+24\xi\lambda-48g_2=0$, or can fall below the phantom divide if the second term in Eq.  (\ref{eq10k}) is negative.  It follows then  that the model (\ref{eq1}) can describe quintessence and phantom phases even without potential. To see this more clearly lets consider the following restriction 
\be\label{eq10l}
2\lambda^3-6\lambda^2+24\xi\lambda-48g_2=0
\ee
which simplifies the Eq. (\ref{eq10i}) and leads to the solution
\be\label{eq10m}
H^2=h_0^2\left[\kappa^2\phi_0^2(\lambda^2-6\xi\lambda+24g_2)+3e^{2\lambda x}\right]^{-1/\lambda}
\ee
where $h_0^2$ is the integration constant. This solution gives the following dark energy EOS
\be\label{eq10n}
w=-1+\frac{2e^{2\lambda x}}{\kappa^2\phi_0^2(\lambda^2-6\xi\lambda+24g_2)+3e^{2\lambda x}}
\ee
lets assume that in the current epoch ($x=0$) the DE EOS parameter takes the value $w_0$. Then setting $x=0$ in (\ref{eq10n}) one finds the following relation
\be\label{eq10p}
\kappa^2\phi_0^2(\lambda^2-6\xi\lambda+24g_2)=-\frac{3w_0+1}{w_0+1}
\ee
By replacing $\xi$ from (\ref{eq10h}) and $g_2$ from the restriction (\ref{eq10l}) one can rewrite the Eq. (\ref{eq10p}) as
\be\label{eq10q}
\frac{2\lambda^3-2\lambda^2}{2\lambda+1}=-\frac{1}{(\kappa\phi_0)^2}\frac{3w_0+1}{w_0+1}
\ee
The r.h.s. of this equation is positive whenever $-1<w_0<-1/3$, i.e. when the current EOS parameter is in the quintessential phase. Corresponding to this, the l.h.s. of the equation is positive if $\lambda>1$ or $\lambda<-1/2$. If the current EOS is $w_0<-1$ or $w_0>-1/3$, then the r.h.s. of eq. (\ref{eq10q}) is negative and the corresponding l.h.s. becomes negative for $\lambda$ in the interval $-1/2<\lambda<1$. We can see the behavior of the initial condition in the EOS parameter depending on the initial conditions on the scalar field $\phi_0$ in Fig. 1.
\begin{figure}[h]
\centering
\includegraphics [scale=0.5]{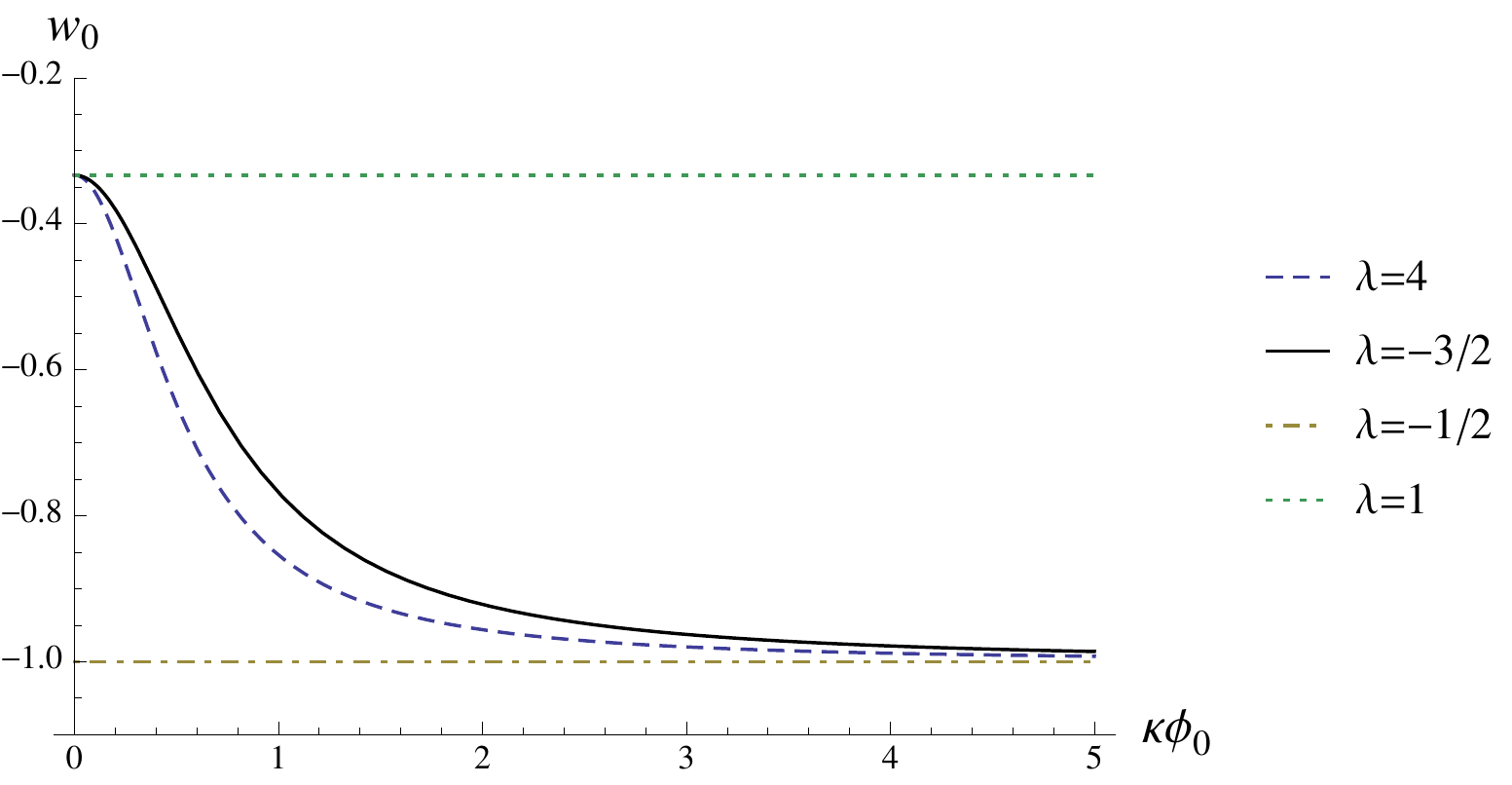}
\caption {\it The initial conditions for the current value of the DE EOS parameter $w_0$ as function of $\kappa\phi_0$ for some values of $\lambda$. The horizontal lines correspond to $w_0=-1/3$ at $\lambda=1$ and the cosmological constant $w_{\Lambda}=-1$ at $\lambda=-1/2$, which are independent of the initial value $\kappa\phi_0$. For $w_0$ in the interval $-1<w_0<-1/3$ the allowed values of $\lambda$ are $\lambda>1$ or $\lambda<-1/2$.}
\end{figure}
In the interval $-1/2<\lambda<1$, where the EOS parameter can take values below the phantom divide, the Hubble parameter is not well defined and therefore the solution (\ref{eq10m}) does not describe phantom DE.
Note that at larger values of $\lambda$ (positive or negative), $w_0$ becomes closer to $-1$.
To agree with current observations on $w_0$, the initial value of the scalar field ($\kappa\phi_0$) should be around 2 or larger. Thus, for $\lambda=4$ and $\kappa\phi=2$ we find $w_0\approx -0.95$. In Fig. 2 we show the evolution of the EOS in terms of the redshift $z$ for initial conditions taken from Fig. 1
\begin{figure}[h]
\centering
\includegraphics [scale=0.5]{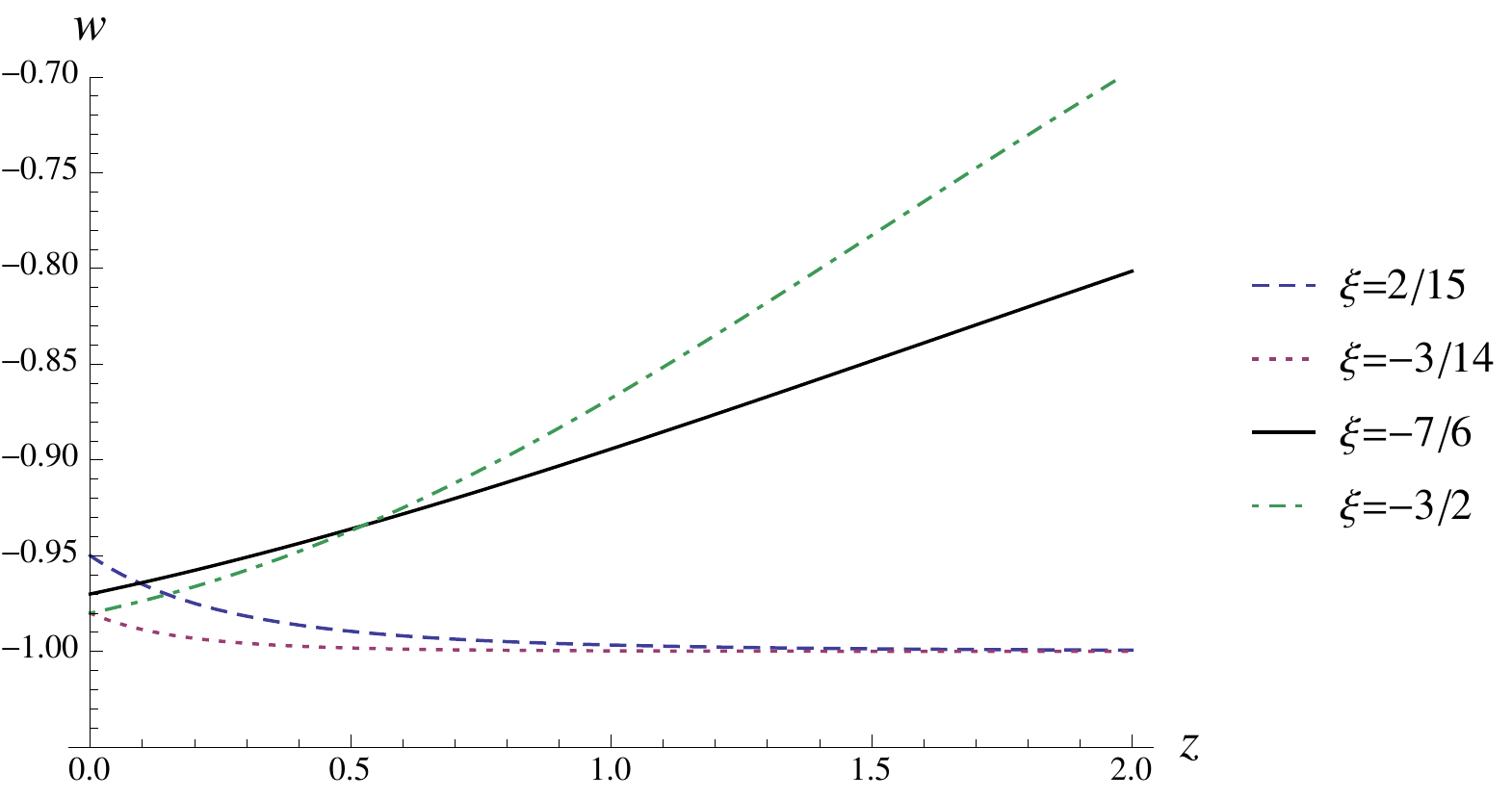}
\caption {\it The evolution of the DE EOS parameter $w$ as function of $z$ for $\xi=(2/15, -3/14, -7/6,-3/2)$ which give ($\lambda=2,3,-1,-3/2$) and $w_0\approx -0,95,-0.98,-0.97,-0.98$ respectively, for $\kappa\phi_0=4$. Though the current values are very similar, the DE evolves to different scenarios depending on $\xi$ (which defines the sign of $\lambda$).}
\end{figure}
The EOS for $\lambda=2,3$ remains very close to $-1$ even for $z>> 2$. In general for positive $\lambda$ the EOS evolves from $-1$ in the past up to $-1/3$ in the far future, and for negative $\lambda$ the EOS evolves from $-1/3$ in the past up to de Sitter state ($-1$) in the distant future. All curves present appropriate quintessence behavior in the "near" past and the present. The behavior of the model depends on the initial value $w_0$ and the non-minimal coupling $\xi$.\\
\noindent Let's consider the general case of the Eq. (\ref{eq10i}) with the solution
\be\label{eq10r}
H^2=h_0^2 e^{-\frac{\gamma}{\beta }x}\left[\kappa^2\phi_0^2\beta+3e^{2\lambda x}\right]^{-\frac{2\beta-\gamma}{2\beta\lambda}}
\ee
where $h_0^2$ is the integration constant and
\[ \beta=\lambda^2-6\xi\lambda+24g_2,\;\;\;\, \gamma=-2\lambda^3+6\lambda^2-24\xi\lambda+48g_2\]
The EOS for this solution is given by
\be\label{eq10s}
w=-1+\frac{\gamma(\kappa\phi_0)^2+6e^{2\lambda x}}{3\beta(\kappa\phi)^2+9e^{2\lambda x}}
\ee
which depending on the sign of $\lambda$ evolves between the values: 
\[ \lim_{x \to -\infty} w=-1+\frac{\gamma}{3\beta}\;\;\;\; {\rm and} \;\;\:\lim_{x \to \infty} w=-\frac{1}{3} \;\;\;{\rm for}\;\;\; \lambda>0 \]
and
\[ \lim_{x \to -\infty}w=-\frac{1}{3} \;\;\;\; {\rm and} \;\;\:\lim_{x \to \infty} w=-1+\frac{\gamma}{3\beta}\;\;\;{\rm for}\;\;\; \lambda<0 \]
It is noted that these limits do not depend on the initial value of the scalar field $\phi_0$. It can be seen from these equations that the solution can contain the phantom phase if one of the parameters $\beta$ or $\gamma$ is negative. Let's consider each case separately. If we assume $\beta<0$ and $ \gamma>0$, then the first $x$-dependent factor in (\ref{eq10r}) increases exponentially with $x$ but the power of the second factor becomes negative, which leads to divergence al some point where $\beta(\kappa\phi_0)^2=-3 e^{2\lambda x}$. So the solutions with $\beta<0$ give rise to Big Rip singularities. These singularities can be avoided if we consider the second case where $\beta>0$ and $\gamma<0$. In this case the second factor in (\ref{eq10r}) is always positive independently of the sign of the power, giving a solution that is always finite with the advantage that can describe a phantom universe. To illustrate the behavior of the solution (\ref{eq10r}), in Fig. 3 we plot the EOS as function of the redshift $z$ for some values of the parameters. 
\begin{figure}[h]
\centering
\includegraphics [scale=0.5]{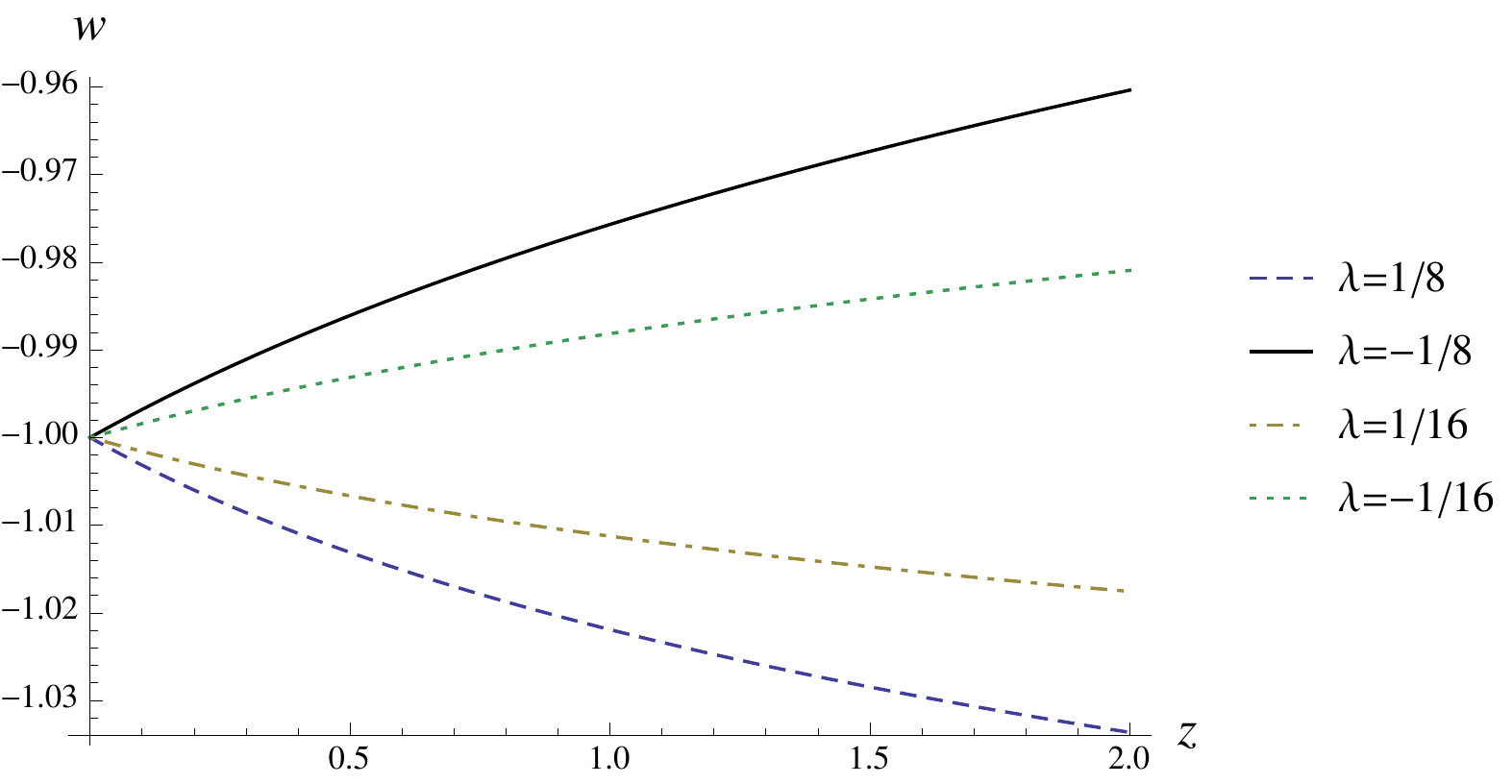}
 \caption {\it The behavior of the DE EOS  $w$ as function of the redshift. The initial value of the scalar field is $\kappa\phi_0=2$ and the current value is $w_0=-1$. For $\lambda=(1/8, -1/8,1/16,-1/16)$ the corresponding coupling parameters take the values $\xi=(5,-5,10,-10)$ and $g_2=0.28$ for all cases.}
\end{figure}
We can see that these curves fit very well to the observed behavior of the DE EOS. According to the curves for $\lambda=1/8,1/16$, the expansion due to DE is currently undergoing the transition from phantom to quintessence phase, and the curves for $\lambda=-1/8,-1/16$ describe un expansion currently going through the transition from quintessence to phantom phase.
There is also possible to have adequate behavior of DE for small values of the coupling $\xi$ as is shown in Figs. 4 and  5, where we show solutions with "almost constant" EOS or with almost constant slope in a wide interval of $z$.
\begin{figure}[h]
\centering
\includegraphics [scale=0.45]{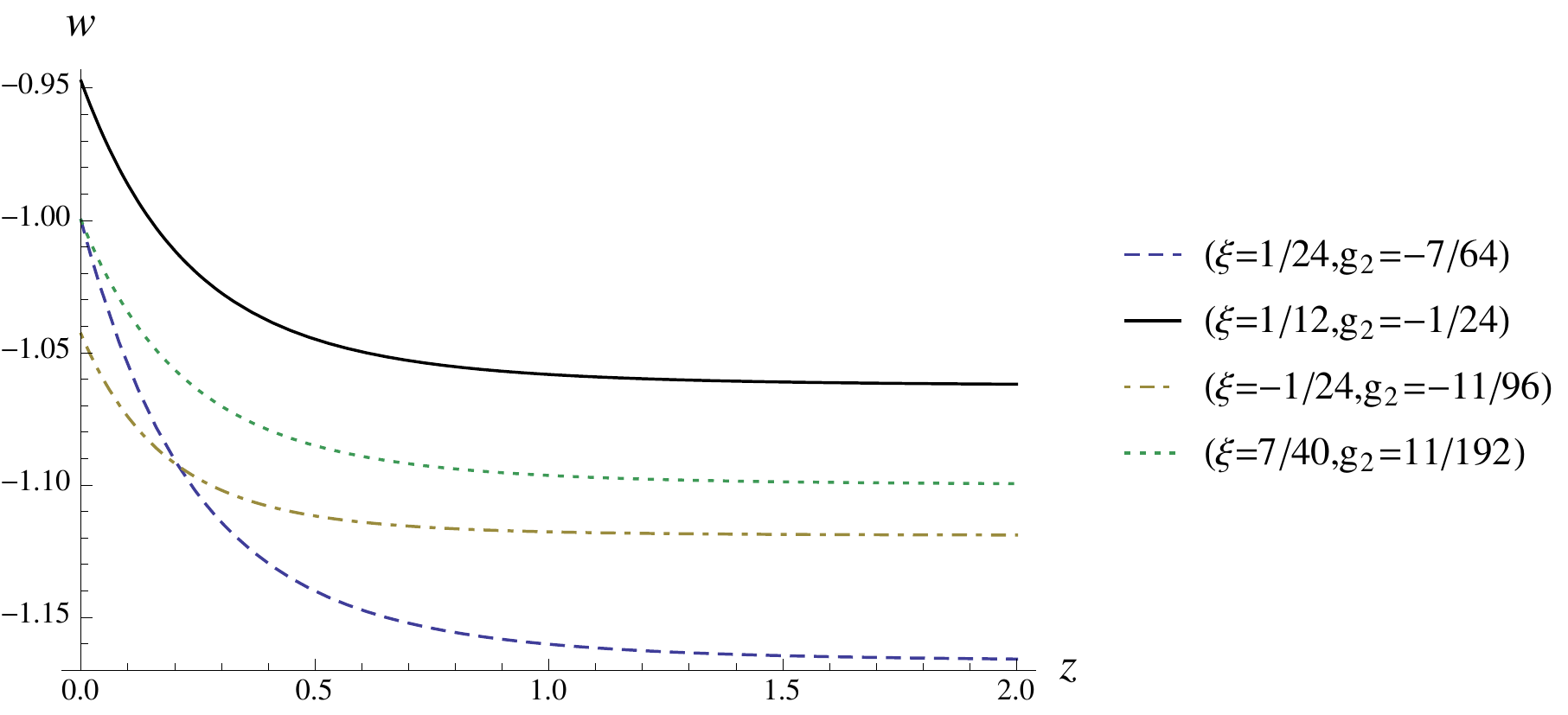}
\caption {\it The DE EOS  $w$ as function of the redshift for small parameters. In all cases $\lambda>0$ and the initial value of the scalar field is $\kappa\phi_0=2$. The curves show different current EOS parameters $w_0$, but all of them are close to $-1$, and present transitions from phantom to quintessence phase in the past ($w_0>-1$), the present ($w_0=-1$) and the future ($w_0<-1$).}
\end{figure}
The EOS shows very little evolution for a wide interval of $z$, and the deviation from almost constant value happens near the present epoch. In fact the model can mimic a behavior very close to the cosmological constant up to recent epochs. 
\begin{figure}[h]
\centering
\includegraphics [scale=0.45]{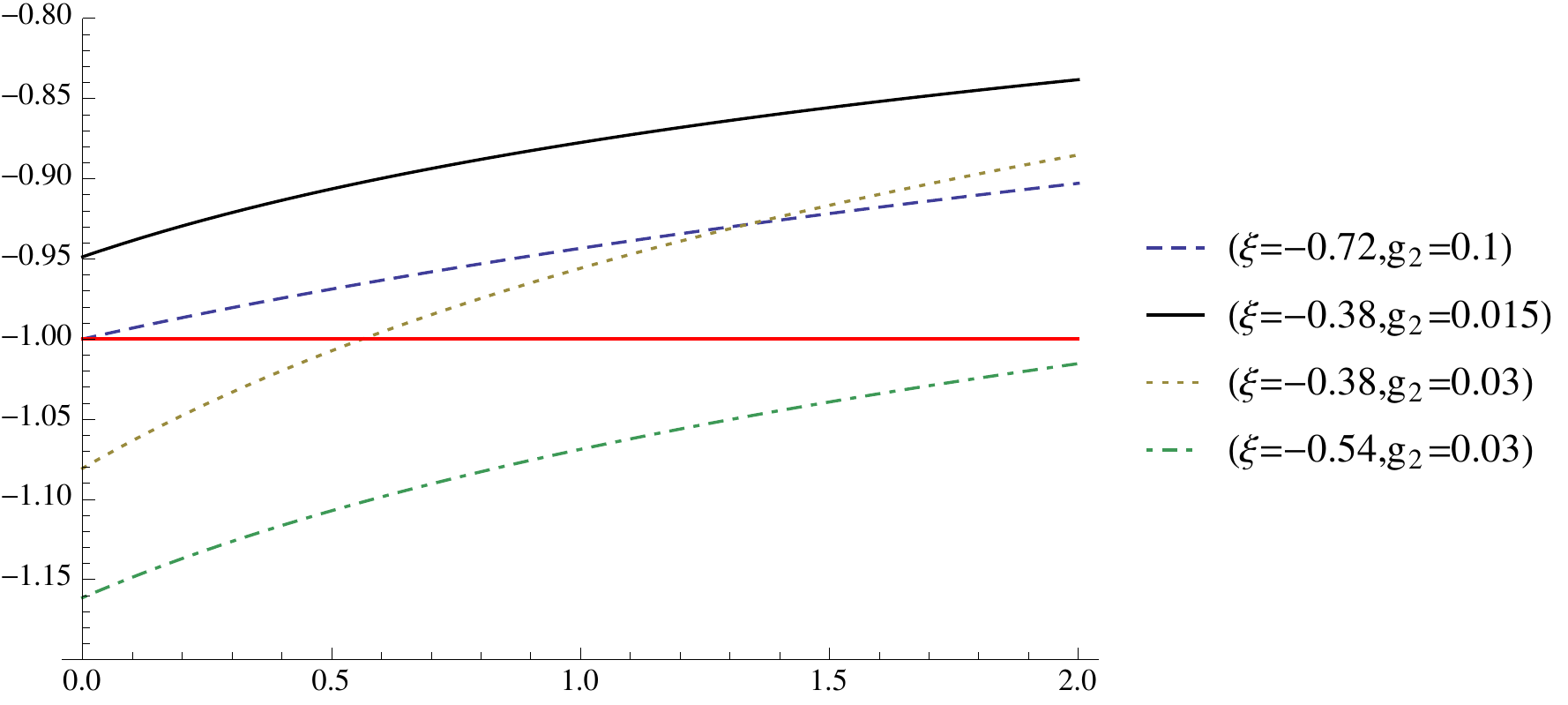}
\caption {Fig. 5 \it The DE EOS  $w$ as function of the redshift for $\lambda<0$. 
In all cases the initial value of the scalar field is $\kappa\phi_0=4$. The curves cover different behaviors of the EOS, but all of them in an acceptable range of values according to the data. Note the transitions from quintessence to phantom phase in the past ($w_0<-1$), the present ($w_0=-1$) and the future ($w_0>-1$). The horizontal line corresponds to the cosmological constant.}
\end{figure}


\section{Reconstructing  quintessence and phantom scenarios}
In an scenario with dominance of DE we assume the scalar field with non-minimal and GB couplings as the  responsible for the energy content of the universe, which is achieved in the model (\ref{eq1}) by setting  ${\cal L}_m=0$ or in the subsequent equations, by making $g_0=0$. In this section we will discuss some important solutions that give appropriate description of late-time universe.  We give explicit examples of reconstructed scalar potential and GB coupling for an accelerated universe in quintessence and phantom phases, with Big Rip and Little Rip singularities. The phantom cosmology can be realized in the present model without resorting to ghost scalar fields.\\
\subsection{Quintessential power-law}

In the FRW background the power-law solutions are of special interest because they represent asymptotic or intermediate states among all possible cosmological evolutions. Let's consider the following functions 

\begin{equation}\label{eq20}
f(\phi)=t_0e^{\phi/\phi_0},\,\,\,\,\,\,\,\, g(t)=p\ln \left(\frac{t}{t_0}\right)
\end{equation}
which according to (\ref{eq15}) lead to

\begin{equation}\label{eq21}
H = \frac{p}{t}, {\kern 1pt} {\kern 1pt} {\kern 1pt} {\kern 1pt} {\kern 1pt} {\kern 1pt} {\kern 1pt} {\kern 1pt} {\kern 1pt} {\kern 1pt} {\kern 1pt} {\kern 1pt} {\kern 1pt} {\kern 1pt} \phi  = {\phi _0}\ln \left( {\frac{t}{{{t_0}}}} \right)
\end{equation}
Replacing (\ref{eq20}) in (\ref{eq16}) one finds the GB coupling as

\begin{equation}\label{eq22}
\eta(\phi)=\frac{t_0^2}{\kappa^2}\left[c_0(p,\xi)+c_1(p,\xi)\phi+c_2(p,\xi)\phi^2\right]e^{2\phi/\phi_0}
\end{equation}
where
$$c_0(p,\xi)=\frac{2p(p+1)^2-(1-\xi+2p+p^2-3p^2\xi)\kappa^2\phi_0^2}{16p^2(1+p)^3},$$
$$c_1(p,\xi)=-\frac{(1+3p)\xi\kappa^2\phi_0}{8p^2(1+p)^2},$$
$$c_2(p,\xi)=-\frac{\xi\kappa^2}{8p(1+p)}$$
and the expression for the potential (\ref{eq17}) becomes

\begin{equation}\label{eq23}
V(\phi)=\frac{1}{\kappa^2 t_0^2}\left[d_0(p,\xi)+d_1(p,\xi)\phi+d_2(p,\xi)\phi^2\right]e^{-2\phi/\phi_0}
\end{equation}
where
\begin{widetext}
\[\begin{gathered}
  {d_0}(p,\xi ) = \frac{{6{p^2}{{(p + 1)}^2}(p - 1) + (5{p^3} + (9 + 24\xi ){p^2} + 3p - 1){\kappa ^2}\phi _0^2}}{{2{{(1 + p)}^3}}}, \\ 
  {d_1}(p,\xi ) = \frac{{12{p^2}\xi {\kappa ^2}{\phi _0}}}{{{{(1 + p)}^2}}}\\
 {d_2}(p,\xi ) =  - \frac{{3(p - 1){p^2}\xi {\kappa ^2}}}{{1 + p}}
\end{gathered} \]
\end{widetext}
The power-law solution can also be realized by the following functions
\begin{equation}\label{eq24}
f(\phi)=t_0\frac{\phi}{\phi_0},\,\,\,\,\, g(t)=p\ln \left(\frac{t}{t_0}\right)
\end{equation}
which lead to the following Gauss-Bonnet coupling

\begin{equation}\label{eq25}
\eta(\phi)=\frac{t_0^2}{8p^2\kappa^2\phi_0^2}\left(\frac{p\phi^2}{p+1}-\frac{(2(2p-1)\xi+1)\kappa^2\phi^4}{4(p-1)}\right)
\end{equation}
and the scalar potential

\begin{eqnarray}\label{eq26}
V(\phi ) =&& \frac{1}{{2({p^2} - 1)t_0^2{\kappa ^2}{\phi ^2}}}\Big[6{p^2}{(p - 1)^2}\phi _0^2\nonumber\\
&& - (p + 1)((6p(p - 3)\xi  - 5)p - 1){\kappa ^2}\phi _0^2{\phi ^2}\Big]
\end{eqnarray}
It can be seen that for large $\phi$ this potential has similar behavior as the potential (\ref{eq25}).
\subsection{Phantom power-law}
Another interesting solution is the phantom power-law, that gives an EoS below the cosmological constant which is not ruled out by current astrophysical observations, but leads to future Big Rip singularity. Assuming the following functions

\begin{equation}\label{eq27}
f(\phi)=t_s-t_0e^{\phi/\phi_0},\,\,\,\,\, g(t)=-p\ln\left(\frac{t_s-t}{t_0}\right)
\end{equation}
which give 
\begin{equation}\label{eq28}
H=\frac{p}{t_s-t},\,\,\,\, \phi=\phi_0\ln\left(\frac{t_s-t}{t_0}\right)
\end{equation}
Replacing in the GB coupling (\ref{eq16}) we find

\begin{equation}\label{eq29}
\eta(\phi)=\left[c_0'(p,\xi)+c_1'(p,\xi)\phi+c_2'(p,\xi)\phi^2\right]e^{2\phi/\phi_0},
\end{equation}
where

$$c_0'(p,\xi)=\frac{2p(p-1)^2-((3\xi-1)p^2+2p+\xi-1)\kappa^2\phi_0^2}{16p^2(p-1)^3}$$

$$c_1'(p,\xi)=\frac{(3p-1)\xi\kappa^2\phi_0}{8p^2(p-1)^2}$$

$$c_2'(p,\xi)=-\frac{\xi\kappa^2}{8p(p-1)}$$
and the potential from (\ref{eq17}) takes the form
\begin{equation}\label{eq30}
V(\phi)=\left[d_0'(p,\xi)+d_1'(p,\xi)\phi+d_2'(p,\xi)\phi^2\right]e^{-2\phi/\phi_0},
\end{equation}
where
\begin{widetext}
\[\begin{gathered}
  {d_0}'(p,\phi ) = \frac{{6{p^2}(p + 1){{(p - 1)}^2} + (5{p^3} - 3(8\xi  + 3){p^2} + 3p + 1){\kappa ^2}\phi _0^2}}{{2{{(p - 1)}^3}}} \\ 
\\
  {d_1}'(p,\xi ) = \frac{{12{p^2}\xi {\kappa ^2}{\phi _0}}}{{{{(p - 1)}^2}}} \\ 
\\
  {d_2}'(p,\xi ) =  - \frac{{3(p + 1){p^2}\xi {\kappa ^2}}}{{p - 1}} \\ 
\end{gathered} \]
\end{widetext}
According to (\ref{eq28}), $\phi\rightarrow -\infty$ at $t\rightarrow t_s$ and the scalar potential (\ref{eq30}) becomes infinity giving rise to the Big Rip singularity. This phantom behavior can also be obtained with the following functions
\begin{equation}\label{eq31}
t=f(\phi)=t_s-t_0\frac{\phi}{\phi_0},\,\,\,\,\, g(t)=-p\ln\left(\frac{t_s-t}{t_0}\right)
\end{equation}
which give the Gauss-Bonnet coupling
\begin{equation}\label{eq32}
\eta(\phi)=-\frac{t_0^2}{8p^2\kappa^2\phi_0^2}\left(\frac{-p\phi^2}{p-1}+\frac{(2(2p+1)\xi-1)\kappa^2\phi^4}{4(p+1)}\right)
\end{equation}
and the scalar potential
\begin{eqnarray}\label{eq33}
V(\phi ) =&& \frac{1}{{2({p^2} - 1)t_0^2{\kappa ^2}{\phi ^2}}}\Big[6{p^2}{(p + 1)^2}\phi _0^2\nonumber\\
&& - (p - 1)((6p(p + 3)\xi  - 5)p + 1){\kappa ^2}\phi _0^2{\phi ^2}\Big]
\end{eqnarray}
This potential also grows toward infinity since $\phi\rightarrow 0$ at $t\rightarrow t_s$.
The expressions for the potentials (\ref{eq25}) and (\ref{eq29})  show a run-away behavior (i.e. $V\rightarrow 0$ at $\phi\rightarrow \infty$) which is  important in the phenomenology of dark energy since it reflects the scaling behavior of the model.
The potentials like (\ref{eq23}) and (\ref{eq30}) which contain combination of exponential and power-law terms can be present in the low energy limit of M-theory as proposed in \cite{andreas, skordis} where it was shown that simple corrections to pure exponential potential produce interesting quintessence accelerating solutions. In the present model (\ref{eq1}) these corrections allow not only quintessence but also phantom scenarios (at least of the power-law type).
Note that from (\ref{eq22}), (\ref{eq23}), (\ref{eq29}), and (\ref{eq30}) follows that when the non-minimal coupling disappears, then only the free term $c_0$, $d_0$, $c_0'$ and $d_0'$  survive in the corresponding expressions, and  the GB coupling and potential become pure exponentials for quintessence and phantom power-law. 
\subsection{The de Sitter solution}
The model (\ref{eq1}) (with ${\cal L}_m=0$) contains de Sitter solution with varying scalar field as we show in the following two examples. Taking the functions
\begin{equation}\label{eq34}
f(\phi)=t_0\frac{\phi}{\phi_0},\,\,\,\,\, g(t)=H_0 t
\end{equation}
that lead to
\begin{equation}\label{eq35}
H=H_0,\,\,\,\,\, \phi=\phi_0\frac{t}{t_0}
\end{equation}
we find the Gauss-Bonnet coupling and the potential as

\begin{equation}\label{eq36}
\begin{gathered}
  \eta (\phi ) =  - \frac{{({\phi _0} + \xi {H_0}{t_0}\phi )\phi }}{{8{t_0}H_0^3}} \\ 
  {\mkern 1mu} V(\phi ) = \frac{{5\phi _0^2}}{{2t_0^2}} + 3H_0^2\left( {\frac{1}{{{\kappa ^2}}} - \xi {\phi ^2}} \right) \\ 
\end{gathered} 
\end{equation}
The following functions also give the de Sitter solution
\begin{equation}\label{eq37}
t=f(\phi)=-t_0\ln\left(\frac{\phi}{\phi_0}\right),\,\,\,\,\, g(t)=H_0 t
\end{equation}
from where $\phi=\phi_0 e^{-t/t_0}$, which lead to the Gauss-Bonnet coupling 
\begin{equation}\label{eq38}
  \eta (\phi ) = \frac{{(1 - 2({H_0}{t_0} + 2)\xi )\phi^2 }}{{16H_0^2({H_0}{t_0} + 2)}}  
\end{equation}
and the expression for the potential (\ref{eq17}) becomes
\begin{equation}\label{eq39}
V(\phi ) = \frac{{3H_0^2}}{{{\kappa ^2}}} - \frac{{((6{H_0}{t_0}({H_0}{t_0} + 2)\xi  - 5){H_0}{t_0} + 2){\phi ^2}}}{{2({H_0}{t_0} + 2)t_0^2}}
\end{equation}
Note that in absence of potential and for the asymptotic case of constant $\phi$ there is not de Sitter solution.

\subsection{Little Rip solution}
This type of solutions represent an alternative to Big Rip \cite{frampton, brevik} where the dark energy density increases with time but without facing a finite time future singularity. In Little Rip solutions  the equation of state parameter $w<-1$ (or $\dot{H}>0$) and causes the same effect of dissociation of matter in the future as the Big Rip solutions, but the scale factor, the density and pressure remain finite at finite time. Let's consider the following scenario
\begin{equation}\label{eq40}
H =H_0 {e^{ht}}{\kern 1pt} {\kern 1pt} {\kern 1pt} {\kern 1pt} {\kern 1pt} {\kern 1pt} {\kern 1pt} {\kern 1pt} {\kern 1pt} {\kern 1pt} {\kern 1pt} {\kern 1pt} {\kern 1pt} {\kern 1pt} {\kern 1pt} {\kern 1pt} {\kern 1pt} {\kern 1pt} {\kern 1pt} {\kern 1pt} {\kern 1pt} {\kern 1pt} {\kern 1pt} {\kern 1pt} {\kern 1pt} {\kern 1pt} {\kern 1pt} {\kern 1pt} \phi =\phi_0 {e^{ht}}
\end{equation}
From where it is seen that $f(\phi ) = \frac{1}{h}\ln \left( {\frac{\phi }{{{\phi _0}}}} \right)$ y $g(t) = \frac{{{H_0}}}{h}{e^{ht}}$. If we replace these functions in  (\ref{eq16}) and  (\ref{eq17}) in absence of matter one obtains

\begin{equation}\label{eq41}
\eta (\phi ) = \frac{{{h^{\text{2}}}(1 - {\text{4}}\xi ){\phi _0}^4}}{{16{H_0}^4{\phi ^2}}} + \frac{{h{H_0}(1 - 4\xi ){\phi _0}^3}}{{8{H_0}^4\phi }} + \frac{{{\phi _0}^{\text{2}}}}{{8{H_0}^2{\kappa ^2}{\phi ^2}}}
\end{equation}
\begin{equation}\label{eq42}
V(\phi ) = {m_1}\phi  + {m_2}{\phi ^2} + {m_3}{\phi ^3} + {m_4}{\phi ^4}
\end{equation}
where
\begin{widetext}
\[{m_1} = \frac{1}{2}\left( {\frac{{6{h^3}(1 - 4\xi ){\phi _0}}}{{{H_0}}} + \frac{{12h{H_0}}}{{{\kappa ^2}{\phi _0}}}} \right),\,\,\,\,{m_2} = \frac{1}{2}\left( {{h^2}(5 - 24\xi ) + \frac{{6{H_0}^2}}{{{\kappa ^2}{\phi _0}^{\text{2}}}}} \right),\,\,\,\,{m_3} =  - \frac{{6h{H_0}\xi }}{{{\phi _0}}},\,\,\,\,{m_4} =  - \frac{{3{H_0}^2\xi }}{{\phi _0^2}}\]
\end{widetext}
\subsection{Solutions with quintessence and phantom phases}
Another interesting scenario that allows the integration in Eqs. (\ref{eq16}) and (\ref{eq17}) presents quintessence and phantom phases and comes from the functions $f(\phi ) = \frac{{{t_0}\phi }}{{{\phi _0}}}$ and $g(t) = p\ln \left( {\frac{t}{{{t_s} - t}}} \right)$, which give
\begin{equation}\label{eq43}
H = p\left( {\frac{1}{t} + \frac{1}{{{t_s} - t}}} \right)\,\,\,\,\,\,\,\,\,\phi  = \frac{{{\phi _0}}}{{{t_0}}}t
\end{equation}
Note that for times $t<t_s/2$ the universe is in the quintessence phase, and enters the phantom phase for times  $t>t_s/2$. The Big Rip singularity occurs at $t=t_s$. The general expressions for $\eta$ and $V$ for arbitrary $p$ depend on hypergeometric functions and are too large. Here we limit ourselves to the special case $p=2$ where the expressions for the GB coupling and potential take the form
\begin{widetext}
\begin{eqnarray}\label{eq44}
\eta (\phi ) =&&  - \frac{1}{{9600{\kappa ^2}{t_s}^{\text{2}}{\phi _0}^{\text{4}}}}\Big[{t_0}^2{\phi ^2}\Big(50{\kappa ^2}(2\xi  - 1){t_0}^2{\phi ^4} + 120{\kappa ^2}(4\xi  + 1){t_0}{t_s}{\phi ^3}{\phi _0}\ln (\phi )\nonumber\\
&& - 8{t_0}{t_s}\phi {\phi _0}\left( {{\text{3}}{\kappa ^{\text{2}}}({\text{4}}\xi {\text{  +  1}}){\phi ^{\text{2}}} - {\text{50}}} \right){\text{  +  25}}{t_s}^{\text{2}}{\phi _0}^{\text{2}}\left( {{\text{3}}{\kappa ^{\text{2}}}({\text{6}}\xi {\text{  +  1}}){\phi ^{\text{2}}} - {\text{8}}} \right)\Big)\Big]
\end{eqnarray}

\begin{eqnarray}\label{eq45}
V\left( \phi  \right) =&&  - \frac{1}{{2{\kappa ^2}{t_0}^2{\phi ^2}{{({t_0}\phi  - {t_s}{\phi _0})}^3}}}\Big[{\phi _0}^{\text{2}}\Big(({\kappa ^{\text{2}}}{t_0}^{\text{3}}{\phi ^{\text{5}}} - {\text{15}}{\kappa ^{\text{2}}}{t_0}^{\text{2}}{t_s}{\phi ^{\text{4}}}{\phi _0} \nonumber\\
&&+ {\text{24}}{\kappa ^{\text{2}}}({\text{4}}\xi {\text{  +  1}}){t_0}{t_s}^{\text{2}}{\phi ^{\text{3}}}{\phi _0}^{\text{2}}\ln (\phi ) + 3{t_0}t{s^2}\phi {\phi _0}^{\text{2}}\left( {{\kappa ^{\text{2}}}({\text{24}}\xi {\text{  +  1}}){\phi ^{\text{2}}}{\text{  +  8}}} \right) \nonumber\\
&&+ {t_s}^{\text{3}}{\phi _0}^{\text{3}}\left( {{\kappa ^{\text{2}}}({\text{24}}\xi {\text{  +  11}}){\phi ^{\text{2}}}{\text{  +  8}}} \right)\Big)\Big]
\end{eqnarray}
\end{widetext}
Note that although the GB coupling (\ref{eq44}) remains finite, the potential (\ref{eq45}) becomes infinite at $t\rightarrow t_s$ ($\phi\rightarrow \phi_0 t_s/t_0$).
\section{Solutions with matter component}
In this case we consider the general equations (\ref{eq16}) and (\ref{eq17}) with $g_0\ne 0$. Below it is shown that even with matter content, the power-law expansion scenarios are solutions. These solutions are important when one of the components is dominating over the other.\\
\subsection{Quintessential solutions}
We consider power-law solutions of the form
\begin{equation}\label{eq46}
H = \frac{p}{t}{\kern 1pt} {\kern 1pt} {\kern 1pt} {\kern 1pt} {\kern 1pt} {\kern 1pt} {\kern 1pt} {\kern 1pt} {\kern 1pt} {\kern 1pt} {\kern 1pt} {\kern 1pt} {\kern 1pt} {\kern 1pt} {\kern 1pt} {\kern 1pt} {\kern 1pt} {\kern 1pt} {\kern 1pt} {\kern 1pt} {\kern 1pt} {\kern 1pt} {\kern 1pt} {\kern 1pt} {\kern 1pt} {\kern 1pt} {\kern 1pt} {\kern 1pt} {\kern 1pt} {\kern 1pt} {\kern 1pt} {\kern 1pt} {\kern 1pt} {\kern 1pt} {\kern 1pt} \phi  = {\phi _0}\ln \left( {\frac{t}{{{t_0}}}} \right)
\end{equation}
which lead to $f(\phi)={t_0}{e^{\phi /{\phi _0}}}$ y $g(t)=p\ln \left( {\frac{t}{{{t_0}}}} \right)$. By replacing in (\ref{eq16}), we find the GB coupling as

\begin{equation}\label{eq47}
\widetilde\eta  = {e^{2\phi /\phi_0}}\left[ {{c_0} + {c_1}\phi   + {c_2}{\phi ^2}}+A{e^{2\phi /\phi_0 }}{e^{ - 3p(1 + w)\phi /\phi_0}} \right]
\end{equation}
where $\widetilde\eta  = \frac{{{\kappa ^2}}}{{{t_0}^2}}\eta $. The constants $c_0$, $c_1$, $c_2$ y $A$ are given by

\[{c_0} = \frac{{2p{{(1 + p)}^2} - {\kappa ^2}{\phi _0}^2(1 - \xi  + 2p + {p^2} - 3{p^2}\xi )}}{{16{p^2}{{(1 + p)}^3}}}\]
\[{c_1} =  - \frac{{\xi {\kappa ^2}{\phi _0}(1 + 3p)}}{{8{p^2}{{(1 + p)}^2}}}\]
\[{c_2} =  - \frac{{\xi {\kappa ^2}}}{{8p(1 + p)}}\]
\[A = \frac{{{g_0}{\kappa ^2}{t_0}^2(w + 1)}}{{8{p^2}(3p(w  + 1) - 4)(p(3w  + 4) - 1)}}\]
By using the functions $f(\phi)={t_0}{e^{\phi /{\phi _0}}}$ y $g(t)=p\ln \left( {\frac{t}{{{t_0}}}} \right)$ in the Eq. (\ref{eq17}) one finds

\begin{equation}\label{eq48}
\widetilde V = {e^{ - 2\phi/\phi_0 }}({d_0} + {d_1}\phi + {d_2}{\phi^2}) + B{e^{ - 3p(1 + w )\phi /\phi_0}}
\end{equation}
where $\widetilde V = {\kappa ^2}{t_0}^2V$ and the constants $d_0$, $d_1$, $d_2$ y $B$ are given by

\begin{widetext}
\[{d_0} = \frac{{6{p^2}( - 1 + p){{(1 + p)}^2} + {\kappa ^2}\phi _0^2( - 1 + p(3 + p(9 + 5p + 24\xi )))}}{{2{{(1 + p)}^3}}},\,\,\,\,{d_1} = \frac{{12\xi {p^2}{\kappa ^2}{\phi _0}}}{{{{(1 + p)}^2}}}\]
\[{d_2} = \frac{{ - 3\xi {p^2}{\kappa ^2}( - 1 + p)}}{{1 + p}},\,\,\,\,\,B = \frac{{{g_0}{\kappa ^2}{t_0}^2(1 - p)}}{{p(3w  + 4) - 1}}\]
\end{widetext}
Note that as the scalar field evolves according to (\ref{eq47}) all the exponential terms in the potential (\ref{eq48}) decrease ($w>-1$) and the potential takes a run-away shape. The scaling behavior of the DE component is clear from the potential  (\ref{eq48}) since the first term is the same potential for the power-law in absence of matter (\ref{eq25})  (which gives $\rho_{\phi}\propto t^{-2}$) and the correction due to the presence of matter also contributes to the density a term $\propto t^{-2}$. 
\subsection{Phantom solutions}
Another interesting cosmology is described by

\begin{equation}\label{eq49}
H = \frac{p}{{{t_s} - t}}{\kern 1pt} {\kern 1pt} {\kern 1pt} {\kern 1pt} {\kern 1pt} {\kern 1pt} {\kern 1pt} {\kern 1pt} {\kern 1pt} {\kern 1pt} {\kern 1pt} {\kern 1pt} {\kern 1pt} {\kern 1pt} {\kern 1pt} {\kern 1pt} {\kern 1pt} {\kern 1pt} {\kern 1pt} {\kern 1pt} {\kern 1pt} {\kern 1pt} {\kern 1pt} {\kern 1pt} {\kern 1pt} {\kern 1pt} {\kern 1pt} {\kern 1pt} {\kern 1pt} {\kern 1pt} {\kern 1pt} {\kern 1pt} {\kern 1pt} \phi  = {\phi _0}\ln \left( {\frac{{{t_s} - t}}{{{t_0}}}} \right)
\end{equation}
From which follows that  $f(\phi ) = {t_s} - {t_0}{e^{\frac{\phi }{{{\phi _0}}}}}$ y $g(t) =  - p\ln \left( {\frac{{{t_s} - t}}{{{t_0}}}} \right)$. This solution describes phantom evolution with $w<-1$ and finite time Big Rip future singularity at $t=t_s$

By replacing the functions (\ref{eq49}) in the GB coupling (\ref{eq16}) it is found that

\begin{equation}\label{eq50}
\widetilde\eta  = {e^{2\phi /{\phi _0}}}({c_0}^\prime  + {c_1}^\prime \phi  + {c_2}^\prime \phi  + A'{e^{2\phi /{\phi _0} }}{e^{3p(1 + w )\phi /{\phi _0} }})
\end{equation}
where

\[{c_0}^\prime  =  - \frac{{{\kappa ^2}{\phi _0}^{\text{2}}(\xi  + p(({\text{3}}\xi  - {\text{1}})p{\text{ + 2}}) - {\text{1}}) - {\text{2}}{{(p - {\text{1}})}^{\text{2}}}p}}{{16{p^2}{{(p - 1)}^3}}}\]

\[{c_1}^\prime  =  - \frac{{{\kappa ^2}{\phi _0}\xi (1 - 3p)}}{{8{p^2}{{(p - 1)}^2}}}\]

\[{c_2}^\prime  =  - \frac{{{\kappa ^2}\xi }}{{8p(p - 1)}}\]

\[A' = \frac{{{g_0}{\kappa ^2}{t_0}^2(w  + 1)}}{{8{p^2}(3p(w  + 1) + 4)(p(3w+ 4) + 1)}}\]
And for the potential (\ref{eq17}) we find the expression

\begin{equation}\label{eq51}
\widetilde V = {e^{ - 2\phi/\phi_0 }}({{d'}_0} + {{d'}_1}\phi  + {{d'}_2}{\phi ^2}) + B'{e^{3p(1 + w )\phi/\phi_0 }}
\end{equation}
where

\begin{widetext}
\[{{d'}_0} = \frac{{6{p^2}(1 + p){{( - 1 + p)}^2} + {\kappa ^2}\phi _0^2(1 + 3p - 9{p^2} + 5{p^3} - 24{p^2}\xi )}}{{2{{( - 1 + p)}^3}}},\,\,\,\,{{d'}_1} = \frac{{12{p^2}{\kappa ^2}\phi_0\xi }}{{{{( - 1 + p)}^2}}}\]
\[{{d'}_2} =  - \frac{{3{p^2}(1 + p){\kappa ^2}\xi }}{{ - 1 + p}},\,\,\,\,\,B' =  - \frac{{{g_0}{\kappa ^2}{t_0}^2(1 + p)}}{{p(3w  + 4) + 1}}\]
\end{widetext}
From (\ref{eq49}) follows that as $t\rightarrow t_s$ the scalar field $\phi$ tends to $-\infty$ and the scalar potential  (\ref{eq51}) undergoes an exponential growth  ($w<-1$), evolving towards the Big Rip singularity at $t\rightarrow t_s$. 
It is interesting to note that even in the presence of matter (minimally coupled as in model (\ref{eq1})) the model may reproduce quintessence or phantom evolution, leading to an effective dark energy universe.
\section{CONCLUSION}
It is well known that different couplings of scalar field to curvature appear in a variety of contexts 
ranging from the quantization of matter fields on curved background to multidimensional theories like string or M-theory. In this regard, it is natural to wonder whether the scalar-tensor couplings predicted by fundamental theories may become important at current low-curvature universe.
In this paper we have considered a model of scalar field with non-minimal and GB couplings in the FRW background, as a model of dark energy. In the model without potential and under appropriate restriction of the GB coupling (given by (\ref{eq10d}) with $g_1=1/(8\kappa^2)$), we have found exact solutions that give appropriate description of 
the accelerated expansion. First under the restriction (\ref{eq10l}) the model describes quintessence dark energy, and its behavior is determined by the initial value $\phi_0$ and the non-minimal coupling $\xi$. Some evolutionary scenarios are shown in Fig. 3, where some of them are very close to the cosmological constant. For the more general case it was shown that the model allows phantom DE evolution without facing Big Rip or any other singularity. We have found that the equation of state has different asymptotic behaviors depending on the couplings of the model, but in all cases is consistent with current observations on DE EOS. Thus for instance, solutions for positive $\lambda=1/8,1/16$ (see Fig. 3) describe an universe with EOS currently $w_0=-1$ and  undergoing the transition from phantom to quintessence phase. But changing the sign, i.e. taking  $\lambda=-1/8,-1/16$ the transition goes in opposite direction, where the universe evolves towards the phantom phase. We can deduce from the equation (\ref{eq10s}) that the current phase ($x=0$) of the EOS depends on the product $\gamma(\kappa\phi_0)^2$: if $\gamma(\kappa\phi_0)^2>-6$, then the EOS is currently $w_0>-1$ (quintessence phase), if $\gamma(\kappa\phi_0)^2<-6$ then $w_0<-1$ (phantom phase). The case depicted in Fig. 3 corresponds to $\gamma(\kappa\phi_0)^2=-6$, where the DE universe is currently passing trough the barrier of the cosmological constant. The model also gives appropriate description of DE for small values of the non-minimal coupling $\xi$ as shown in Figs. 4 and 5, which describe scenarios with phantom phase, with transition from phantom to quintessence phase (Fig. 4) and from quintessence to phantom phase (Fig. 5) in different cosmological epochs. In all cases we have found that the model covers a variety of DE scenarios, all of them in the range of the measurements made by the different collaborations.\\
\noindent We also considered an approach that allows in principle to reconstruct the potential an the GB coupling for arbitrary given cosmological scenarios \cite{sergei5}. We showed explicit examples of reconstruction for quintessence and phantom power-law solutions, where the obtained potentials are of exponential form with polynomial factor. These type of potentials were proposed in \cite{andreas, skordis} as coming from low-energy limit of higher dimensional theories (after compactification of extra dimensions), and which present attractive features for description of realistic cosmologies. The fact that the potential was reconstructed from power-law solutions, indicates that if we start from the model (\ref{eq1}) with GB coupling and potential given by (\ref{eq22}) and (\ref{eq23}) respectively, then the  dynamics of this model on the FRW background gives rise to scaling solutions which play an important role in quintessence scenarios \cite{copeland}. The polynomial factor in the potential and GB coupling disappears in absence of the non-minimal coupling ($\xi=0$) as follows from expressions (\ref{eq22}), (\ref{eq23}), (\ref{eq29}) and (\ref{eq30}), leading to pure exponential GB couplings and potentials, which are typical of string-inspired gravity models.\\
\noindent The model could also be reconstructed for a Little Rip solution where the Hubble parameter increases exponentially and the scalar potential becomes a fourth-order polynomial. A quintom scenario with quintessence and phantom phases was also reconstructed for the Hubble parameter (\ref{eq43}), which led to the GB coupling (\ref{eq44}) and potential (\ref{eq45}). It is worthwhile to emphasize that the phantom scenario could be realized without introducing the ghost scalar field, which is quite attractive for a viable model of dark energy. For the de Sitter solution we have found that both the potential and the GB coupling depend on the square of the field. \\
\noindent Resuming, the non-minimal and GB couplings allow exact solutions that give rise to interesting cosmological scenarios describing the DE in a manner consistent with current astrophysical observations. These solutions include phantom expansion without singularities and transitions between quintessence and phantom phases. It was also demonstrated that in principle any given late-time cosmological solution relevant to current observations (including quintessence,  phantom, quintom, de Sitter, Little Rip) may be reconstructed by scalar-tensor model with GB and non-minimal coupling. The reconstructed potential for power-law quintessence solutions has been considered previously as coming from higher dimensional gravity theories. All these facts enable us to conclude that the combination of GB and non-minimal couplings gives a good explanation of the dynamics of the late time universe, and could play an important role in revealing the nature of the DE.
\begin{acknowledgments}
This work was supported by Universidad del Valle under project CI 7987, DFJ acknowledges support from COLCIENCIAS, Colombia.
\end{acknowledgments}

\nocite{*}

\end{document}